%% file: main.tex
\documentclass[twoside, twocolumn,
 prl,
 floats,
 amsmath,amssymb,
 aps,
]{revtex4-1}

\usepackage[english]{babel}
\usepackage[paperwidth=210mm,paperheight=297mm,centering,hmargin=2cm,vmargin=2.5cm]{geometry}
\usepackage{graphicx}
\usepackage[colorlinks=true, allcolors=blue]{hyperref}
\usepackage{changes}
\usepackage{setspace}

\begin{document}


\title{High-harmonic generation in semi-Dirac and Weyl semimetals with broken time-reversal symmetry: Exploring merging of Weyl nodes}%

\author{Luka Medic}
 \email{luka.medic@ijs.si}
\author{Jernej Mravlje}%
\author{Anton Ram\v{s}ak}%
\author{Toma\v{z} Rejec}%
\affiliation{Jo\v{z}ef Stefan Institute, Jamova 39, SI-1000 Ljubljana, Slovenia}
\affiliation{Faculty of Mathematics and Physics, University of Ljubljana, Jadranska	19, SI-1000 Ljubljana, Slovenia}
 

\date{\today}

\begin{abstract}
We explore anomalous high-harmonic generation in a model that realizes a transition from a broken time-reversal symmetry Weyl-semimetal to a semi-Dirac regime, i.e. a gapless semimetal with dispersion that is parabolic in one direction and conical in the other two. We point out the intensity of the induced anomalous high harmonics is high in the semi-Dirac regime. For Weyl semimetals, we reveal anomalous high harmonics are due to excitations at momenta where the dispersion is not strictly linear and that in the linearized low-energy theory the anomalous response is harmonic only. Our findings aid experimental characterization of Weyl, Dirac, and semi-Dirac semimetals.
\end{abstract}

\maketitle

\begin{spacing}{1.02}

The high harmonic generation (HHG) in condensed matter systems has attracted significant attention due to its potential applications in ultrafast optics \citep{manzoni2015ultrafast, huttner2017ultrahigh, weber2017similar, sie2019ultrafast, nematollahi2019weyl, seo2023completely, gao2020chiral, boland2023narrowband} and as a characterization method \citep{cheng2020, reinhoffer2022high, zong2023emerging, takasan2021current, goulielmakis2022high, li2020attosecond, murakami2022anomalous, lv2021high}. HHG was proposed to be efficient in Dirac (DSMs) and Weyl semimetals (WSMs)  as they are gapless materials with linear energy dispersion \citep{mikhailov2007non, mikhailov2008nonlinear, lv2021high, wang2022highly, hafez2018extremely, kovalev2020non, dantas2021nonperturbative, germanskiy2022ellipticity, avetissian2012creation} and exhibit very high carrier mobilities \citep{shekhar2015extremely, liang2015ultrahigh, narayanan2015linear, kumar2017extremely, kaneta2022high}. Here we focus on the anomalous optical response (i.e. in the current in a direction perpendicular to the electric field) \citep{avetissian2022high, bharti2022high, wilhelm2021semiconductor}. In WSMs, the linear anomalous response is proportional to the separation of the Weyl nodes and thus related to topology in these materials. Anomalous high harmonics were proposed as a means for probing the Berry curvature \citep{silva2019topological, luu2018measurement, liu2017high, yue2023characterizing}. In WSMs, the Berry curvature diverges at the Weyl points \citep{lv2021high, nathan2022topological} and one might expect this enhances the generation of anomalous high harmonics (although the contribution due to Berry curvature may not be dominant for all driving frequencies \citep{yue2023characterizing}).

\begin{figure}
\centering
\includegraphics[width=0.485\textwidth]{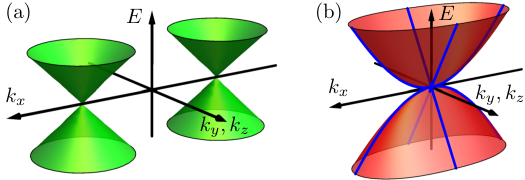}
\caption{Low energy spectra in (a) Weyl and (b) semi-Dirac semimetals (blue lines highlight the parabolic-2D-conical dispersion).}
\label{fig:dispersion}
\end{figure}

In WSMs, the band touchings are characterized by their topological charge which determines the chirality of massless Weyl fermions \citep{armitage2018weyl} -- and by combining two Weyl fermions with opposite chiralities one obtains a Dirac fermion. However, the merging of a pair of Weyl nodes may also give rise to a topologically trivial yet unconventional semi-Dirac regime with parabolic dispersion in the direction of the Weyl nodes separation vector and linear dispersion in the other two directions (i.e. parabolic-2D-conical; see Fig. \ref{fig:dispersion}) \citep{mohanta2021semi, wang2023correlated}. Note that the semi-Dirac regime is distinct from the double-Weyl regime \citep{fang2012multi, armitage2018weyl, bharti2023role} that arises from merging the Weyl nodes with the same chiralities and has parabolic dispersion in two directions. An example of a material that realizes a semi-Dirac regime is $\textrm{SrNbO}_3$ where the fourfold degenerate semi-Dirac point is protected by a nonsymmorphic symmetry \citep{mohanta2021semi}. Another candidate with such parabolic-2D-conical dispersion, albeit with a tiny energy gap of $6\,\textrm{meV}$, is $\text{ZrTe}_5$ \citep{manzoni2015ultrafast, seo2023completely, martino2019two, cottin2020probing, rukelj2020distinguishing, wang2020examining, wang2021gapped, wang2022unconventional, wang2023quantum, wang2023correlated}.

HHG has already been studied in DSMs and WSMs with broken time-reversal (TRS) \citep{wilhelm2021semiconductor, avetissian2022high, bharti2022high} or inversion symmetry (IS)~\citep{liu2017high, luu2018measurement, li2022high}. Despite great interest, the qualitative understanding of anomalous HHG (AHHG) in WSMs seems poor, and what aspects of Weyl physics are really contributing is unclear. Here we disentangle the contributions to the response that can be described in terms of well-separated Weyl nodes with conical dispersion~\citep{cano2017chiral, vazifeh2013electromagnetic}, which was argued to lead to AHHG response that increases with the distance between the nodes~\citep{avetissian2022high}, from the contributions that come from non-linearity of dispersion, which becomes large when the nodes approach, and to investigate the effects of the merging of a pair of Weyl nodes to a Dirac or semi-Dirac node.

We consider a minimal 3D model, which describes a transition between two-node WSMs with broken TRS and semi-DSMs with a parabolic-2D-conical energy dispersion. We use the semiconductor Bloch equations (SBE) \cite{floss2018ab, wilhelm2021semiconductor, avetissian2022high, mrudul2021high} to explore the dynamics of the system under the influence of an infrared pulse. We investigate how the anomalous response varies with the separation between the Weyl nodes. In a regime of large Weyl-node separation, we find that whereas the linear response is large, the higher harmonic intensity drops and becomes negligible. Our results reveal the key aspect of AHHG in WSMs: (i) the response {\it vanishes} for linearized Weyl dispersion, and hence, (ii) the response arises from {\it deviations from strict linearity}. The response becomes large when the Weyl nodes merge and a semi-DSM is realized. We also consider the effects of tilting the Weyl cones. Very recently, a related study has found enhancement of AHHG in multi-Weyl systems \citep{bharti2023role}.


\textit{Model and methods ---} We begin with the Hamiltonian \cite{hasan2017discovery, armitage2018weyl, bharti2022high} $H(\mathbf{k}) = \sum_i d_i(\mathbf{k}) \sigma_i$ where $\mathbf{k}=(k_x,k_y,k_z)$ represents the crystal momentum,  and Pauli matrices $\sigma_i$ act on the pseudospin degree of freedom, such as orbital or sublattice. The components of the vector $\mathbf{d}(\mathbf{k})$ are given by $d_x(\mathbf{k}) = t_x \cos(a k_x) + t_y \cos(b k_y) + t_z \cos(c k_z) - \gamma$, $d_y(\mathbf{k}) = t_y\sin(b k_y)$ and $d_z(\mathbf{k}) = t_z\sin(c k_z)$. The Hamiltonian has orthorhombic symmetry, but for simplicity, we will investigate it for a symmetric choice of parameters $t_x=t_y=t_z=t$ and $a=b=c$ and take dimensionless units $a=1$, $t=1$. 

With varying $\gamma$, the Hamiltonian exhibits three distinct regimes, as detailed in the Supplemental Material (SM) S1 \citep{medic2023supplemental}: (a) a trivial insulator phase occurs for $|\gamma| > 3$, (b) a WSM phase with one pair of Weyl nodes for $1 < |\gamma| < 3$, and two pairs of Weyl nodes for $|\gamma| < 1$, and (c)  at the critical points, $|\gamma| = 3$ and $|\gamma| = 1$, a semi-Dirac regime arises, featuring one band touching and three band touchings, respectively. In the semi-Dirac regime, the band touchings exhibit a parabolic dispersion along the separation vector of the Weyl nodes ($k_x$ direction), while in the perpendicular plane ($k_y$--$k_z$) they display a 2D conical dispersion. The energy dispersions for the conduction ($c$) and valence ($v$) bands are given by $\epsilon_{c,v}(\mathbf{k}) = \pm | \mathbf{d}(\mathbf{k}) |$.

The Hamiltonian lacks TRS but possesses IS and mirror symmetry, expressed by $\sigma_x H(-\mathbf{k}) \sigma_x = H(\mathbf{k})$ and $H(k_x, k_y, k_z) = H(-k_x, k_y, k_z)$, respectively. Additionally, Hamiltonians with opposite values of $\gamma$ are related by
\begin{equation}
\label{eq:symGamma}
    \sigma_z H(\mathbf{R} - \tilde{\mathbf{k}}, -\gamma) \sigma_z = H(\mathbf{k}, \gamma),
\end{equation}
where we employed the notation $\tilde{\mathbf{k}} = (k_x, -k_y, k_z)$ and $\mathbf{R}=(\pi,\pi,\pi)$.

At time $t=0$ we expose the system to an ultrashort laser pulse with a duration $T$.
Electric field of the laser pulse is described by $\mathbf{E}(t) = E_0 f_\tau\left(t-T/2\right) \sin(\omega_0 t) \hat{\mathbf{e}}$ where $E_0$ is the electric field amplitude, $\hat{\mathbf{e}}$ is the polarization vector (in our calculations we use $\hat{\mathbf{e}} = \hat{e}_z$), $\omega_0$ is the carrier frequency and $f_\tau(t)$ is the Gaussian envelope function $f_\tau(t) = \exp\left[-t^2/(2\tau^2)\right]$ where $\tau$ characterizes the pulse width.

The dynamics of the system are described by the SBE for density matrix elements $\rho_{mn}^{\mathbf{k}}$ in the Houston basis \cite{nematollahi2019weyl, houston1940acceleration, krieger1986time}:
\begin{align}
\label{eq:SBE}
    \partial_t \rho_{mn}^{\mathbf{k}}(t) = -i \omega_{mn}^{\mathbf{k}(t)}\rho_{mn}^{\mathbf{k}}(t) - \frac{1-\delta_{mn}}{T_2}\rho_{mn}^{\mathbf{k}}(t) \nonumber \\
    + i \mathbf{E}(t) \cdot \sum_l \left[ \mathbf{d}_{ln}^{\mathbf{k}(t)}\rho_{ml}^{\mathbf{k}} - \mathbf{d}_{ml}^{\mathbf{k}(t)}\rho_{ln}^{\mathbf{k}} \right]
\end{align}
where we work in units $e_0=\hbar=1$, $\mathbf{k}(t) = \mathbf{k} + \mathbf{A}(t)$, $\mathbf{A}(t) = -\int_0^t \mathbf{E}(t') \mathrm{d}t'$ is the vector potential, $\omega_{mn}^{\mathbf{k}} = \epsilon_m(\mathbf{k}) - \epsilon_n(\mathbf{k})$ specifies the energy gap, $T_2$ is (effective) decoherence time \cite{floss2018ab, kilen2020propagation, wilhelm2021semiconductor} and $\mathbf{d}_{mn}^{\mathbf{k}} = \langle m,\mathbf{k} | i \partial_\mathbf{k}  | n, \mathbf{k} \rangle$ are transition dipole moments. Initial condition is $\rho_{mn}(t=0) = \delta_{mv}\delta_{nv}$, i.e. fully occupied valence band.

The current density is
\begin{equation}
\label{eq:totalCurrent}
    \mathbf{J}(t) = \sum_{\mathbf{k}} \mathbf{J}(\mathbf{k},t) = \sum_{\mathbf{k}} \left(-\sum_{m,n} \rho_{mn}^{\mathbf{k}} \mathbf{p}_{nm}^{\mathbf{k}(t)}\right)
\end{equation}
where $\mathbf{p}_{nm}^{\mathbf{k}} = \langle n,\mathbf{k} | \partial_\mathbf{k} H(\mathbf{k})  | m, \mathbf{k} \rangle$ are group-velocity matrix elements. Finally, using the Larmor's formula \cite{floss2018ab}, we obtain the spectrum which is proportional to the intensity of emitted radiation:
\begin{equation}
    \mathcal{I}(\omega) = \left| \mathcal{FT}\left[ \frac{\mathrm{d}}{\mathrm{d}t} \mathbf{J}(t) \right] \right|^2 = \omega^2 | \mathbf{J}(\omega) |^2.
\end{equation}
To obtain a time series of a specific harmonic with $\omega = m \omega_0$ ($m \in \mathbb{N}$), we use the following filter:
\begin{equation}
\label{eq:Jfilter}
    \mathbf{J}^\omega(t) = \mathcal{FT}^{-1} \left[ \int_{\omega - \delta\omega}^{\omega + \delta\omega} \mathcal{FT} \left[ \left( \mathbf{J}(t) \right) \right]  \mathrm{d}\omega' \right],
\end{equation}
where the width of the filter is $2\delta\omega = \omega_0$.

We report results for the following values of parameters (taking $a=1$ \AA~and $t=1$ eV): $\hbar\omega_0=0.2$ eV, $E_0 = 0.043$ V/\AA, $\tau = 40$ fs, $T = 790$ fs, $T_2 = 3$ fs. Our choice of parameters follows Ref. \citep{bharti2022high}. We used the discretization of the BZ with $60^3$ points.

\begin{figure*}
\centering
\includegraphics[width=1.\textwidth]{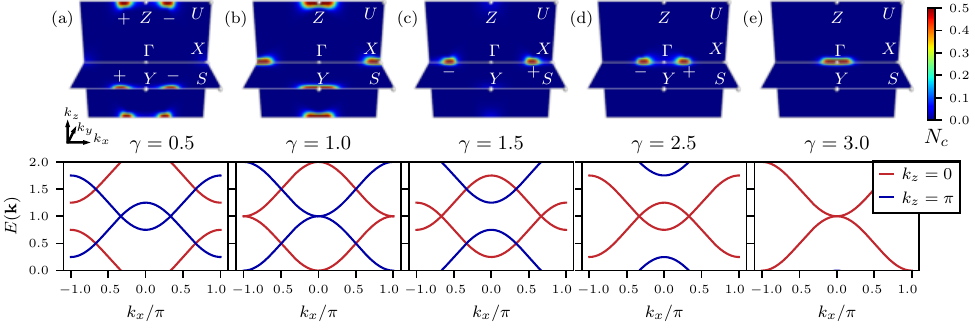}
\caption{\textit{Top:} Excitations $N_c$ at final time $T$ for different $\gamma$ (increasing from left to right). Excitations occur near band touchings due to the divergence of transition dipole moments. Positive (negative) topological charges are indicated by $+$ ($-$). \textit{Bottom:} The energy bands are depicted in red and blue colors for momenta along the high symmetry lines $\Gamma$--$X$ ($k_y=k_z=0$) and $Z$--$U$ ($k_y=0$, $k_z=\pi$), respectively. For $|\gamma| < 1$ there are two pairs of Weyl cones (a); for $\gamma = 1 $ there are three semi-Dirac points (b); for $1 < \gamma < 3$ there are two Weyl cones (c,d); and when $\gamma = 3$ there is one semi-Dirac point (e). Diagrams are similar for negative $\gamma$ where the roles of $\Gamma$ and $\mathrm{R}$ points of the primitive orthorhombic lattice are reversed which is a consequence of the symmetry (\ref{eq:symGamma}).}
\label{fig:excitations}
\end{figure*}

\textit{Results ---} Excitations at the final time $N_c(\mathbf{k}, T) = \rho_{cc}(\mathbf{k}, T)$, i.e. after the pulse has passed, are shown in Fig. \ref{fig:excitations} (top). The bottom panels depict dispersions along two high symmetry lines where energy band touchings occur. We see that higher concentrations of excitations coincide with the band touchings where transition dipole moments $\mathbf{d}_{vc}$ diverge \cite{avetissian2022high}.

\begin{figure}
\centering
\includegraphics[width=0.485\textwidth]{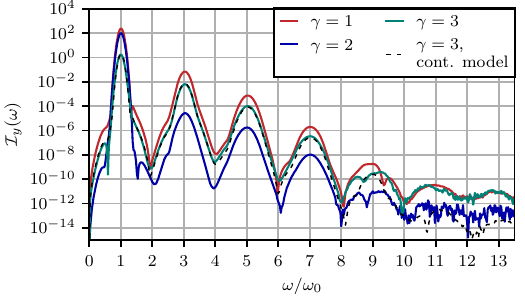}
\caption{Anomalous spectrum $\mathcal{I}_y$ for $\gamma = 1$, $2$ and $3$ (corresponding to 3 semi-Dirac nodes, 1 pair of Weyl nodes, and 1 semi-Dirac node, respectively). The dashed line illustrates the continuous model, with the current density integrated over crystal momenta near the $\Gamma$ point within the $\left[-\pi/3,\pi/3\right]^3$ range.}
\label{fig:spectrum}
\end{figure}

In Fig. \ref{fig:spectrum}, we present the $\mathcal{I}_y$ component, perpendicular to both the electric field polarization vector and the Weyl node separation vector, for three different values of $\gamma$. We observe that the even harmonics ($\omega = 2n \omega_0$) vanish due to the IS \cite{boyd2008nonlinear, bharti2022high}. While the spectra may appear similar at first glance, it is important to note that the intensity of higher-order harmonics in the semi-Dirac regimes ($\gamma=1$ and $\gamma=3$) is significantly higher compared to the WSM regime ($\gamma = 2$).

To study the dynamics of the semi-Dirac regime in the low-energy limit, i.e. in the case of low frequency and strength of the laser pulse, we compare the lattice result for $\gamma = 3$ with the corresponding continuous model obtained by Taylor expansion around the $\Gamma$ point (see SM S1 \citep{medic2023supplemental}). The results, shown in Fig. \ref{fig:spectrum}, demonstrate agreement between the two models, confirming that the dominant contributions arise from the region near $\Gamma$ where excitations are present.

\begin{figure}
\centering
\includegraphics[width=0.485\textwidth]{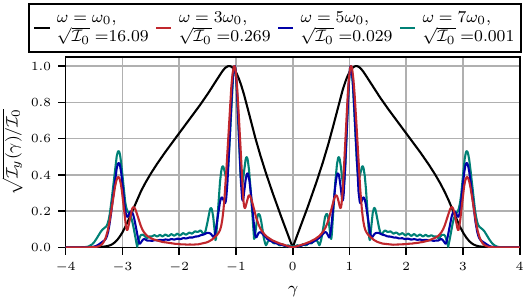}
\caption{Peak intensities for different $\gamma$ and anomalous harmonics at odd multiples of $\omega_0$. For each $\omega$, intensities $\mathcal{I}_y(\gamma)$ are scaled by a factor $\mathcal{I}_0 = \max_{\gamma} \!\left(\mathcal{I}_y\right)$.}
\label{fig:HHGgamma}
\end{figure}

From the spectra obtained for different values of $\gamma$ we extract the peak intensities of the odd harmonics, $\omega = (2n+1)\omega_0$, and present them in Fig. \ref{fig:HHGgamma}. Intensities show a symmetry $\mathcal{I}_y(\gamma)=\mathcal{I}_y(-\gamma)$ and also $\mathcal{I}_y(\gamma=0)=0$ holds. These observations are a result of IS and symmetry Eq. (\ref{eq:symGamma}) (for the derivation see SM S2 \citep{medic2023supplemental}). Moreover, we observe a qualitative difference between the first harmonic and higher-order harmonics, where the former increases from $\gamma=0$ to $\gamma=1$ and then decreases from $\gamma=1$ to $\gamma=3$. In contrast, the higher-order harmonics exhibit distinct peaks at $|\gamma|=1$ and $3$ (the reasons underlying the fine structure seen in these peaks are discussed below). The comparatively higher peaks at $\gamma=3$, in comparison to $\gamma=1$, are due to the co-occurrence of three semi-Dirac points, which is related to the symmetric choice $t_x=t_y=t_z$. If hoppings were unequal, the semi-Dirac points would occur at different values of $\gamma$ (SM S1 \citep{medic2023supplemental}), and the present peaks at $\gamma=1$ would split.

\begin{figure}[h]
\centering
\includegraphics[width=1.\textwidth]{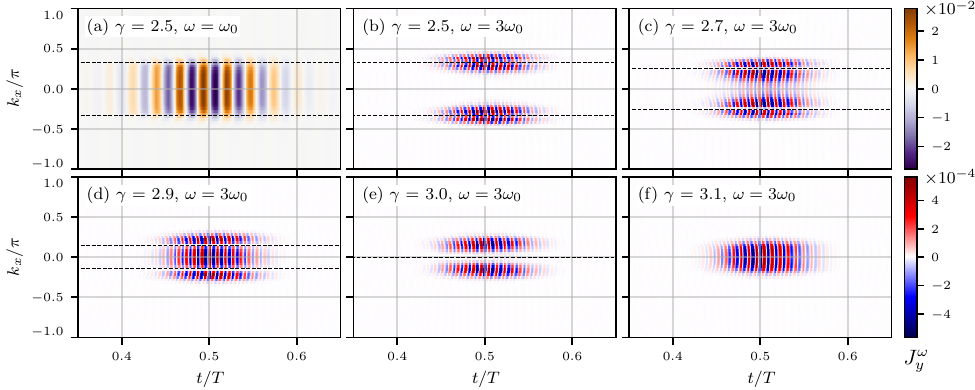}
\caption{Time series of sliced contributions to the anomalous current ${J}_y^{\omega}(t,k_x) = \sum_{k_y,k_z} {J}_y^{\omega}(t,\mathbf{k})$ for (a) the first harmonic, $\omega = \omega_0$, and (b--f) the third harmonic, $\omega = 3 \omega_0$, at various $\gamma$ values. Dashed lines indicate the positions of Weyl nodes (a--d) and a semi-Dirac node (e).}
\label{fig:anomalousCurrentSlices}
\end{figure}

To analyze the results presented in Fig. \ref{fig:HHGgamma}, it is beneficial to illustrate the anomalous response decomposed into its harmonic components $J_y^{\omega}(t)$ using Eq. (\ref{eq:Jfilter}). Figure \ref{fig:anomalousCurrentSlices} displays the outcomes of this decomposition for the first (a) and the third (b--f) harmonic, where we summed the contributions over planes ($k_y$--$k_z$) perpendicular to the Weyl node separation vector. We will refer to the results of Fig. \ref{fig:anomalousCurrentSlices} in the subsequent analysis of AHHG.

The dominant contribution to the first harmonic can be explained by the linear intraband anomalous Hall current \cite{wilhelm2021semiconductor, avetissian2022high, vampa2014theoretical} which is proportional to the separation between the pairs of Weyl nodes (see Fig. \ref{fig:anomalousCurrentSlices}a) $J_y^{\omega_0}(t) \sim \frac{1}{2}\sum_k (\mathbf{E} \times \mathcal{B}^{\mathbf{k}(t)})_y = \frac{1}{2} E_z(t) \frac{1}{(2\pi)^2}\int_{-\pi}^\pi C_x(k_x) \mathrm{d}k_x = -\frac{k_0}{4\pi^2} E_z(t) $
where $\mathcal{B}^{\mathbf{k}} = 2i \mathbf{d}_{cv}^{\mathbf{k}} \times \mathbf{d}_{vc}^{\mathbf{k}}$ is the Berry curvature, $C_x(k_x)= \frac{1}{2\pi}\iint (\mathcal{B}^\mathbf{k})_x \mathrm{d}k_y\mathrm{d}k_z$ is Chern number for 2D slice at $k_x$ and $2 k_0 = 2\arccos(\gamma-2)$ is the separation between Weyl nodes for $1 < \gamma < 3$. When integrating the Chern number over $k_x$ we took into account $C_x(|k_x|<k_0) = -1$ and $0$ otherwise. Similar reasoning applies for $-1 < \gamma < 1$, however, in this case, there are two pairs of Weyl nodes (see Fig. \ref{fig:excitations}a) which contribute $C_x(|k_x|<k_0) = 2$. As a result, in the range $-1<\gamma<1$, the intensity $\mathcal{I}_y(\omega=\omega_0)$ increases twice as fast as it decreases for $1<\gamma<3$ \cite{avetissian2022high}. Trivially, the linear anomalous response also goes to zero upon the merger of a pair of Weyl nodes to a Dirac node.

Understanding the different qualitative behavior of anomalous higher-order harmonics as the Weyl nodes merge, specifically the pronounced peaks at $|\gamma|=1$ and $|\gamma|=3$ in Fig. \ref{fig:HHGgamma}, requires recognizing that high harmonics result from the presence of excitations ($N_c$) and interband polarization ($\rho_{vc}$) \cite{avetissian2022high}. Consequently, only regions around the Weyl or semi-Dirac points contribute significantly to higher-order harmonics, which implies the response cannot be expressed in terms of  $C_x(k_x)$. Furthermore, we have evaluated separately different contributions to the response \cite{avetissian2022high} and found that the Berry curvature contribution and the interband contribution to the high harmonics are of similar magnitude which precludes interpretation of the signal in terms of the Berry curvature alone.

To closely examine the higher-order response, we first consider the case with a pair of well-separated Weyl nodes \citep{footnoteWellseparated} (Fig. \ref{fig:anomalousCurrentSlices}b). For each well-separated Weyl node at $\mathbf{k}_0$, we can take the region where dispersion is conical and make a Taylor expansion in terms of $\mathbf{q} = \mathbf{k} - \mathbf{k_0}$. Then, as detailed in SM S4 \citep{medic2023supplemental}, we have a $C_{2z}$ rotational symmetry with respect to each Weyl node, where the axis is defined by the electric field polarization vector: $\sigma_z H(-q_x, -q_y, q_z) \sigma_z = H(q_x, q_y, q_z)$, which leads to the relation for the anomalous current:
\begin{equation}
\label{eq:symWeylNode}
    {J}_y(q_x, q_y, q_z) = -{J}_y(-q_x, -q_y, q_z).
\end{equation}
As a result, contributions to the anomalous response near each Weyl node in the linearized limit cancel out (generalization to 3D Dirac cones applies trivially). Hence, the only non-vanishing contribution to AHHG may arise from regions far enough from Weyl nodes, where the linear dispersion approximation no longer applies, but still have non-zero excitations; in other words, a finite response is a result of deviations from strict linearity near each Weyl node \footnote{In a different setting, unrelated to HHG, the importance of deviations from strict linearity was discussed in Ref. \citep{bharti2023massless}.}. Interestingly, this result applies (under some limitations) even for tilted type-I WSMs (for more details, see SM S5 \citep{medic2023supplemental}). Note that our conclusion regarding the vanishing of anomalous high-harmonic response in the regime of well-separated Weyl cones differs from the one presented in Ref. \cite{avetissian2022high}. The resolution of this discrepancy is provided in SM S6 \citep{medic2023supplemental}.

Now, let us consider the case when a pair of Weyl nodes start to approach each other and eventually merge into a semi-Dirac point. Utilizing mirror symmetry $H(k_x,k_y,k_z)=H(-k_x,k_y,k_z)$ (as detailed in SM S7 \citep{medic2023supplemental}), we find
\begin{equation}
\label{eq:mirrorSym}
    {J}_y(k_x,k_y,k_z) = {J}_y(-k_x,k_y,k_z).
\end{equation}
Thus, studying only half-space $k_x > 0$ is sufficient. Bringing a Weyl cone in the proximity of the $\Gamma$ point, e.g. as $\gamma \to 3$, brings along two important changes. Firstly, the density of excitations near $\Gamma$ point starts to increase which leads to increasing contributions to the anomalous current that do not have a counterpart that would cancel them since near $\Gamma$ linear approximation does not hold (Fig. \ref{fig:anomalousCurrentSlices}c); consequently, the total anomalous current is not vanishing and increases as the Weyl node moves towards $\Gamma$. Secondly, as the Weyl node keeps approaching the $\Gamma$ point, the region near $\Gamma$ with non-linear dispersion and non-zero excitations starts to shrink (Fig. \ref{fig:anomalousCurrentSlices}d) which first leads to a decrease in the intensity of the anomalous high harmonics; but then as the region near $\Gamma$ keeps shrinking the outer region contributions ($k_x > k_0$) start to dominate and HHG intensity increases again. This explains the dip near $\gamma=2.9$ observed in Fig. \ref{fig:HHGgamma}. In the limit $\gamma \to 3$ (Fig. \ref{fig:anomalousCurrentSlices}e) only outer regions ($k_x > k_0 \to 0$) contribute to HHG and since ${J}_y(q_x=0) = 0$ for each Weyl node we have in this limit ${J}_y(k_x=0) \to 0$. Hence, in the semi-Dirac regime ($\gamma = 3$) we still have ${J}_y(k_x) = {J}_y(-k_x)$ (from Eq. \ref{eq:mirrorSym}) with additional constraint ${J}_y(k_x=0) = 0$. When the parameter $\gamma$ is increased beyond $3$, the high-harmonic response continues to increase because regions near $k_x=0$ also begin to contribute to the overall anomalous current (see Fig. \ref{fig:anomalousCurrentSlices}f). However, as the gap widens further, the density of excitations quickly decreases, resulting in an overall reduced response in the $\mathcal{I}_y$.

We considered a system at half-filling, where the Fermi surface resides at the Weyl points. With small dopings, the validity of our results persists, holding true for both the WSM and semi-Dirac regimes. This is because the regions in the BZ where states in both bands are either empty or occupied do not contribute to the current. For high dopings, the pronounced peaks in Fig. \ref{fig:HHGgamma} would start to diminish, as regions near $k_x=0$ in Fig. \ref{fig:anomalousCurrentSlices}e would cease to contribute to the AHHG.


\vspace{.3cm}

\textit{Discussion --} 
In conclusion, our study clarifies the distinction between AHHG in WSMs, 3D DSMs, and semi-DSMs. AHHG arises from deviations from the conical dispersion, regardless of the specific (Weyl or Dirac) semimetallic system under consideration. Conversely, in the semi-Dirac regime, the dispersion is quadratic along the Weyl node separation vector, leading to enhanced AHHG. As a result, this distinction provides a valuable opportunity for experimental differentiation between the semi-Dirac and 3D Dirac regimes.

\clearpage
\onecolumngrid

\twocolumngrid

We reveal a symmetry that leads to the vanishing of the AHHG in the well-separated Weyl regime and explain enhancement of AHHG (which is maximized near the semi-Dirac regime) in terms of departure from that idealized Weyl situation thereby providing a complementary insight from that offered in a study of multi-Weyl systems  \citep{bharti2023role} that stressed the role of the higher population in the conduction band.

We used a simplified lattice model to study AHHG. In future research, it would be interesting to conduct realistic calculations on $\textrm{SrNbO}_3$ \citep{mohanta2021semi} and $\textrm{ZrTe}_5$ \citep{martino2019two, cottin2020probing, rukelj2020distinguishing}. Notably, $\textrm{SrNbO}_3$ might be worth investigating, as it hosts symmetry-protected semi-Dirac points at the Brillouin-zone boundary, and applying an external magnetic field generates pairs of Weyl nodes. To generalize our results, it is important to consider the fourfold degeneracy exhibited in such materials. Additionally, investigating similar effects in 2D materials like black phosphorus \cite{kim2015observation}, which exhibit a mixed quadratic and linear dispersion profile at critical points, would be of interest as well.

It would also be instructive to explore the significance of the over-tilted, type-II Weyl cones on AHHG in our TRS-broken model, which we leave for further studies.

\vspace{0.3cm}


We thank G. Mkrtchian for useful correspondence.

We acknowledge support from the Slovenian Research and Innovation Agency (ARIS) under Contract No. P1-0044; AR was also supported by Grant J2-2514. JM acknowledges support by ARIS under Grant No. J1-2457.

\end{spacing}

\bibliographystyle{apsrev4-1}
\input{bibOutput.bbl}

\end{document}


\preprint{???}

\title{Supplemental Material\\[0.5cm]High-harmonic generation in semi-Dirac and Weyl semimetals with broken time-reversal symmetry: Exploring merging of Weyl nodes}%

\author{Luka Medic}
\author{Jernej Mravlje}%
\author{Anton Ram\v{s}ak}%
\author{Toma\v{z} Rejec}%
\affiliation{Jo\v{z}ef Stefan Institute, Jamova 39, SI-1000 Ljubljana, Slovenia}
\affiliation{Faculty of Mathematics and Physics, University of Ljubljana, Jadranska	19, SI-1000 Ljubljana, Slovenia}

\maketitle


\onecolumngrid

\appendix
\section{S1. Weyl and semi-Dirac regimes}
\label{appendix:i}

\begin{figure}[h]
\centering
\includegraphics[width=1.\textwidth]{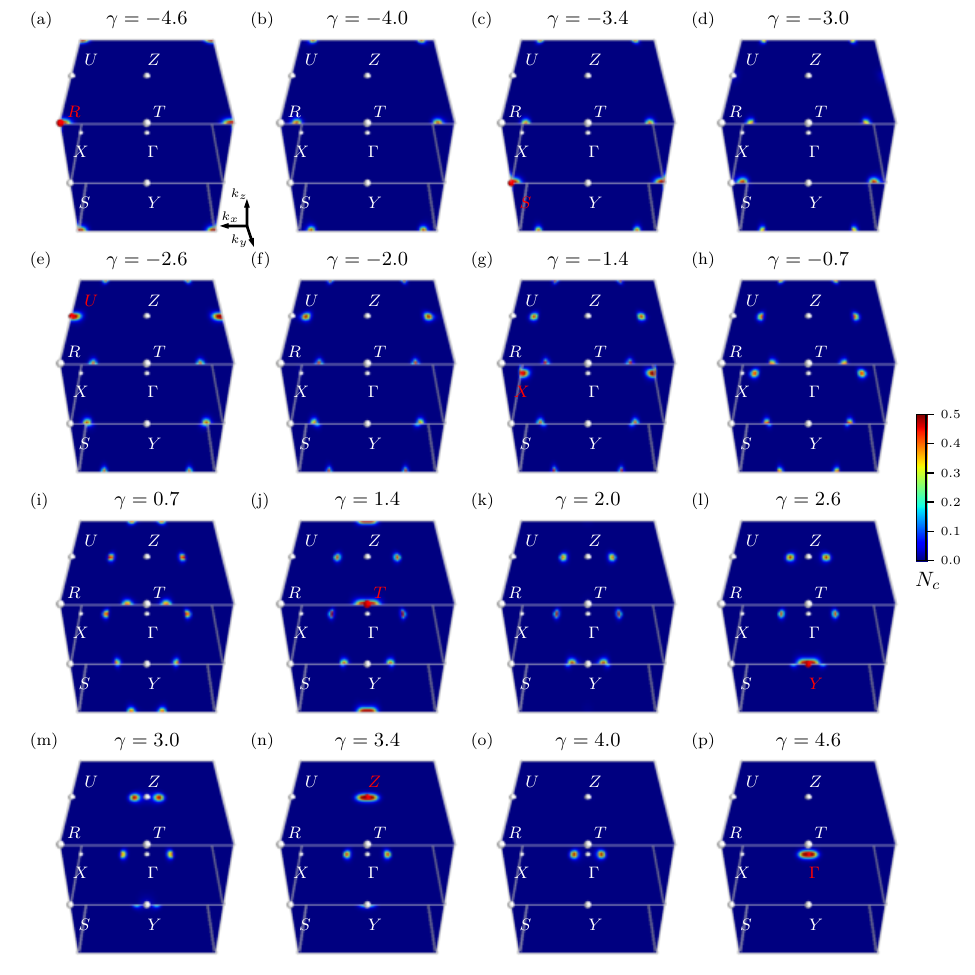}
\caption{Excitations $N_c$ in the BZ of a primitive orthorhombic lattice with lattice constants $a=b=c=1$ and hopping parameters $t_x = 3$, $t_y=1$, and $t_z = 0.6$. Semi-Dirac points (8 in total) are highlighted with red labels (a,c,e,g,j,l,n,p).}
\label{fig:excitationsSupp}
\end{figure}

To determine the nodes of the model Hamiltonian described in the main text we search for solutions of $\epsilon_{c,v} = 0$, or equivalently $d_x = d_y = d_z = 0$. From $d_y = d_z = 0$, we observe that $b k_y$ and $c k_z$ should take values $0$ or $\pi$. Considering the constraint $d_x = 0$, we obtain $|\cos(a k_x)| = |(\gamma \mp t_y \mp t_z) / t_x | \leq 1$, where the upper and lower signs correspond to $b k_y, c k_z = 0$ and $\pi$, respectively. The nodes thus occur for values of $\gamma$ that obey  $-t_x \pm t_y \pm t_z \leq \gamma \leq t_x \pm t_y \pm t_z$. For values of $\gamma$ in the interior of that interval one finds pairs of Weyl-nodes, whereas at extremal points ($\gamma = \pm t_x \pm t_y \pm t_z$) the Weyl-nodes merge resulting in semi-Dirac points (8 total).  These semi-Dirac points are located at time-reversal invariant momenta (TRIM). This is illustrated in Fig. \ref{fig:excitationsSupp}, where the regions of higher excitation densities $N_c$ indicate the positions of Weyl and semi-Dirac nodes in the Brillouin zone (BZ).

Let us examine the low-energy spectrum. We denote the positions of the four pairs of Weyl nodes as $\pm \mathbf{k}_0^{(\pm, \pm)}$, where the $+$ and $-$ signs in the superscript correspond to $b k_y, c k_z = 0$ and $\pi$, respectively. The nodes are separated in the $k_x$ direction. Let $2 k_0^{(\pm, \pm)}$ (or $2 k_0$ for brevity) denote the distance between them. We expand energy $\epsilon^{(\pm, \pm)}(\mathbf{q}) \equiv \epsilon(\mathbf{k}_0^{(\pm, \pm)}+\mathbf{q})$ up to the second order in the displacement vector $\mathbf{q}$:
\begin{align}
    \epsilon^{(\pm, \pm)}(\mathbf{q}) \approx \sqrt{ \left( \frac{(v_x q_x)^2}{2 t_x} \cos(a k_0) + v_x q_x \sin(a k_0)\right)^2 + \left( v_y q_y \right)^2 + \left(v_z q_z\right)^2 } \tag{S1.1}
\end{align}
where we introduced $v_x = a t_x$, $v_y = b t_y$, and $v_z = c t_z$. For $\sin(a k_0) = 0$, we have a semi-Dirac point at $k_x = 0$ or $\pi/a$ with a parabolic ($q_x^2$) and 2D conical (in the $q_y$--$q_z$ plane) dispersion. For $\sin(a k_0) \neq 0$, we have a pair of Weyl nodes at $k_x = \pm k_0$ with, in general, anisotropic linear dispersion that depends on $k_0$.

\begin{figure}[h]
\centering
\includegraphics[width=1.\textwidth]{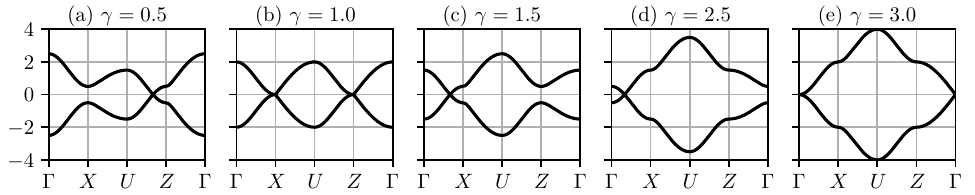}
\caption{Energy spectra along high symmetry lines for various values of $\gamma$. The parabolic-2D-conical dispersion is observed in (b) at $X$ and $Z$, and in (e) at $\Gamma$.}
\label{fig:dispersionSymmetryLines}
\end{figure}

The above derivation simplifies in the specific case we present in the paper, where $a = b = c = 1$ and $t_x = t_y = t_z = 1$. In this case, we identify five qualitatively different regimes:
\begin{itemize}
    \item $|\gamma| > 3$: This regime corresponds to a gapped insulator phase, similar to the one presumably observed in the $\textrm{ZrTe}_5$.
    \item $|\gamma| = 3$: In this regime, we have a semi-DSM with a single semi-Dirac node. For $\gamma = 3$ it is at $\mathbf{k}_0=(0,0,0)$ and for $\gamma = -3$ it is at $\mathbf{k}_0=(\pi,\pi,\pi)$.
    \item $1 < |\gamma| < 3$: Within this regime, we encounter WSMs with one pair of Weyl nodes; for $1< \gamma < 3$ they are at $\mathbf{k}_0=(\pm\arccos(\gamma - 2),0,0)$ and for $-3 < \gamma < -1$ they are at $\mathbf{k}_0=(\pm\arccos(\gamma + 2),\pi,\pi)$
    \item $|\gamma| = 1$:  This regime represents a semi-DSM with three semi-Dirac nodes. For $\gamma = 1$ they are at $\mathbf{k}_0=(\pi,0,0)$, $(0,\pi,0)$ and $(0,0,\pi)$, while for $\gamma = -1$ they are at $\mathbf{k}_0=(0,\pi,\pi)$, $(\pi,0,\pi)$ and $(\pi,\pi,0)$.
    \item $|\gamma| < 1$: This regime corresponds to a WSM featuring two pairs of Weyl nodes which are at $\mathbf{k}_0=(\pm\arccos(\gamma),\pi,0)$ and $(\pm\arccos(\gamma),0,\pi)$.
\end{itemize}
Conical dispersion around the Weyl nodes becomes isotropic when $k_0 = \pi/2$ ($\gamma=\pm 2$). In Fig. \ref{fig:dispersionSymmetryLines}, an additional chart, accompanying Fig. 2 of the main text, illustrates the energy spectra along high symmetry lines.

\newpage
\clearpage

\section{S2. Derivation of symmetry relation $\mathcal{I}(\gamma) = \mathcal{I}(-\gamma)$}
\label{appendix:ii}

In this section we will use the notation $\tilde{\mathbf{v}} = (v_x, -v_y, v_z)$ where $\mathbf{v}$ is an arbitrary 3D vector.

From the symmetry relation stated in Eq. (1) of the main text, we obtain the following relationships:
\begin{align}
\label{eq:apx2:eps}
    \omega_{mn}(\mathbf{k}, -\gamma) &= \omega_{mn}(\mathbf{R} - \tilde{\mathbf{k}}, \gamma) \tag{S2.1} \\
\label{eq:apx2:d}
    \mathbf{d}_{mn}(\mathbf{k}, -\gamma) &= - \tilde{\mathbf{d}}_{mn}(\mathbf{R} - \tilde{\mathbf{k}}, \gamma) e^{i\varphi_{mn}} \tag{S2.2} \\
\label{eq:apx2:p}
    \mathbf{p}_{mn}(\mathbf{k}, -\gamma) &= - \tilde{\mathbf{p}}_{mn}(\mathbf{R} - \tilde{\mathbf{k}}, \gamma) e^{i\varphi_{mn}} \tag{S2.3}
\end{align}
where the factor $e^{i\varphi_{mn}}$ represents an arbitrary gauge factor.

By inserting identities (\ref{eq:apx2:eps})--(\ref{eq:apx2:p}) into the SBE [i.e., Eq. (2) of the main text], we obtain:
\begin{equation}
\label{eq:apx2:rho}
    \rho_{mn}^\mathbf{k}(t, -\gamma, \mathbf{E}) = \rho_{mn}^{\mathbf{R} - \tilde{\mathbf{k}}}(t, \gamma, -\tilde{\mathbf{E}}) e^{-i\varphi_{mn}}. \tag{S2.4}
\end{equation}
Inserting the relations (\ref{eq:apx2:p}) and (\ref{eq:apx2:rho}) into Eq. (3) of the main text, we obtain a relation for the total current:
\begin{equation}
    \mathbf{J}\left(-\gamma, \mathbf{E}\right) = -\tilde{\mathbf{J}}\left(\gamma, -\tilde{\mathbf{E}}\right), \tag{S2.5}
\end{equation}
which is gauge independent.

Additionally, by following similar steps as described above, we can establish that inversion symmetry leads to the reversal of the total current when the direction of the electric field is reversed, i.e. $\mathbf{E} \to - \mathbf{E}$ results in $\mathbf{J} \to -\mathbf{J}$.

Combining these two results, we obtain:
\begin{equation}
    \mathbf{J}\left(-\gamma, \mathbf{E}\right) = \tilde{\mathbf{J}}\left(\gamma, \tilde{\mathbf{E}}\right). \tag{S2.6}
\end{equation}

In the specific case presented in our paper, where the driving pulse has $z$ polarization, we have $\mathcal{I}(\gamma) = \mathcal{I}(-\gamma)$, and $\mathcal{I}_y(\gamma=0) = 0$.

\section{S3. Dependence of response on $\omega_0$ and $E_0$;  BZ maps of contributions to the response}
\label{appendix:iii}

\begin{figure}
\centering
\includegraphics[width=.95\textwidth]{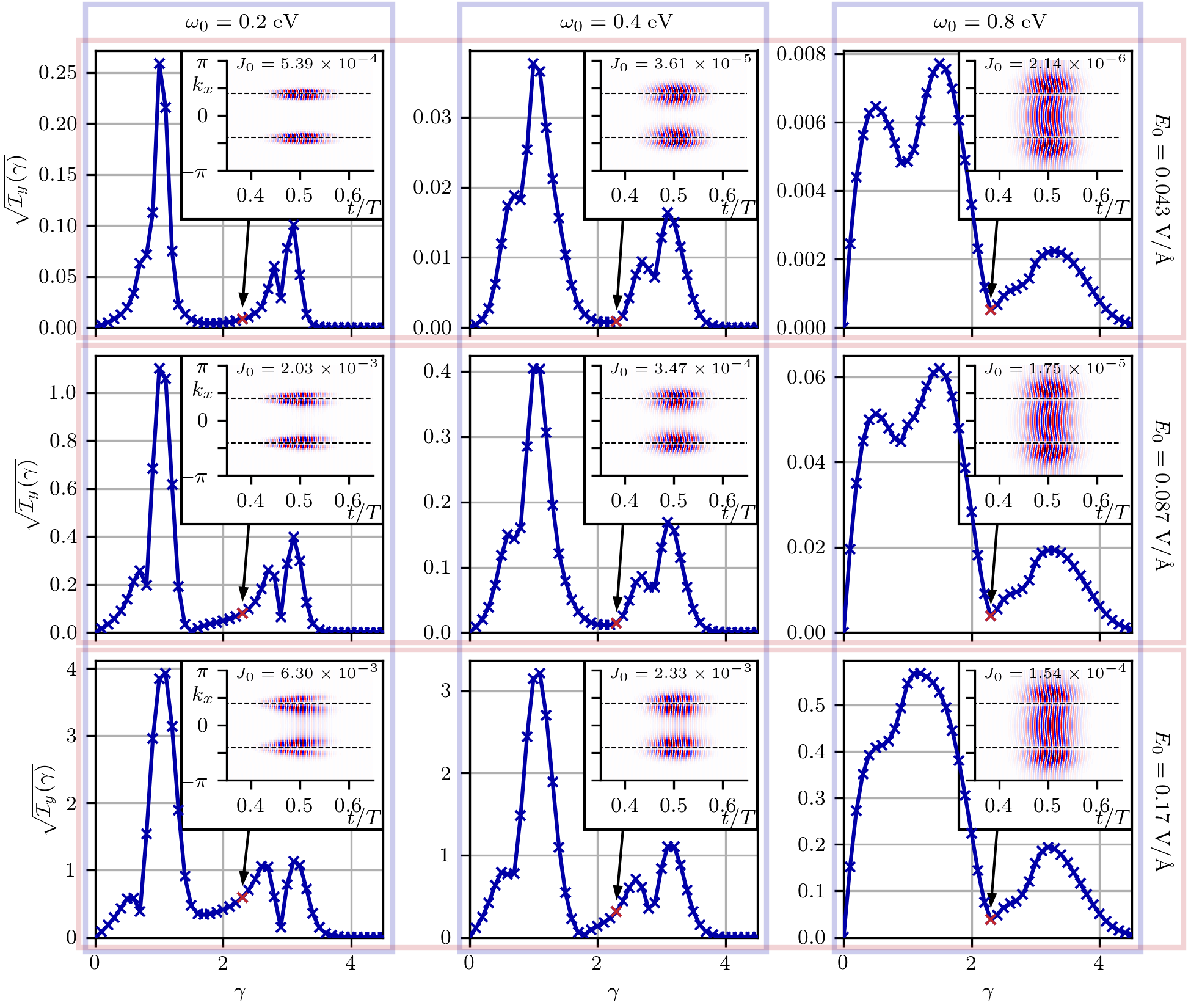}
\caption{Peak intensities of the third anomalous harmonic, $\mathcal{I}_y(\omega=3\omega_0)$, as a function of $\gamma$, for various electric field carrier frequencies $\omega_0$ (columns) and amplitudes $E_0$ (rows). \textit{Inset:} Time series of sliced contributions to the third harmonic of the anomalous current ${J}_y^{3\omega_0}(t,k_x) = \sum_{k_y,k_z} J_y^{3\omega_0}(t,\mathbf{k})$. $J_0$ indicates the maximal amplitude of ${J}_y^{3\omega_0}(t,k_x)$ for a given $\omega_0$, $E_0$, and $\gamma=2.3$. Dashed lines indicate the positions of Weyl nodes.}
\label{fig:figHHGgammaDifferentE}
\end{figure}

In Fig.~\ref{fig:figHHGgammaDifferentE} we show peak intensities of third anomalous harmonic as a function of $\gamma$ for several pulse frequencies $\omega_0$ and amplitudes of the electric field $E_0$. For small $\omega_0$ and $E_0$ one clearly sees response sharply peaking in the proximity of the two semi-Dirac regimes, $\gamma = 1$ and $\gamma = 3$. In the intermediate Weyl regimes, the response is suppressed, which is explained in terms of the cancelation of the response for each individual Weyl point as long as it can be considered well-separated. The localization of the response next to the Weyl points for small $\omega_0$ and $E_0$ is illustrated in the insets that show the third harmonic contribution to anomalous current as a function of time and $k_x$. 

\begin{figure}
\centering
\includegraphics[width=.95\textwidth]{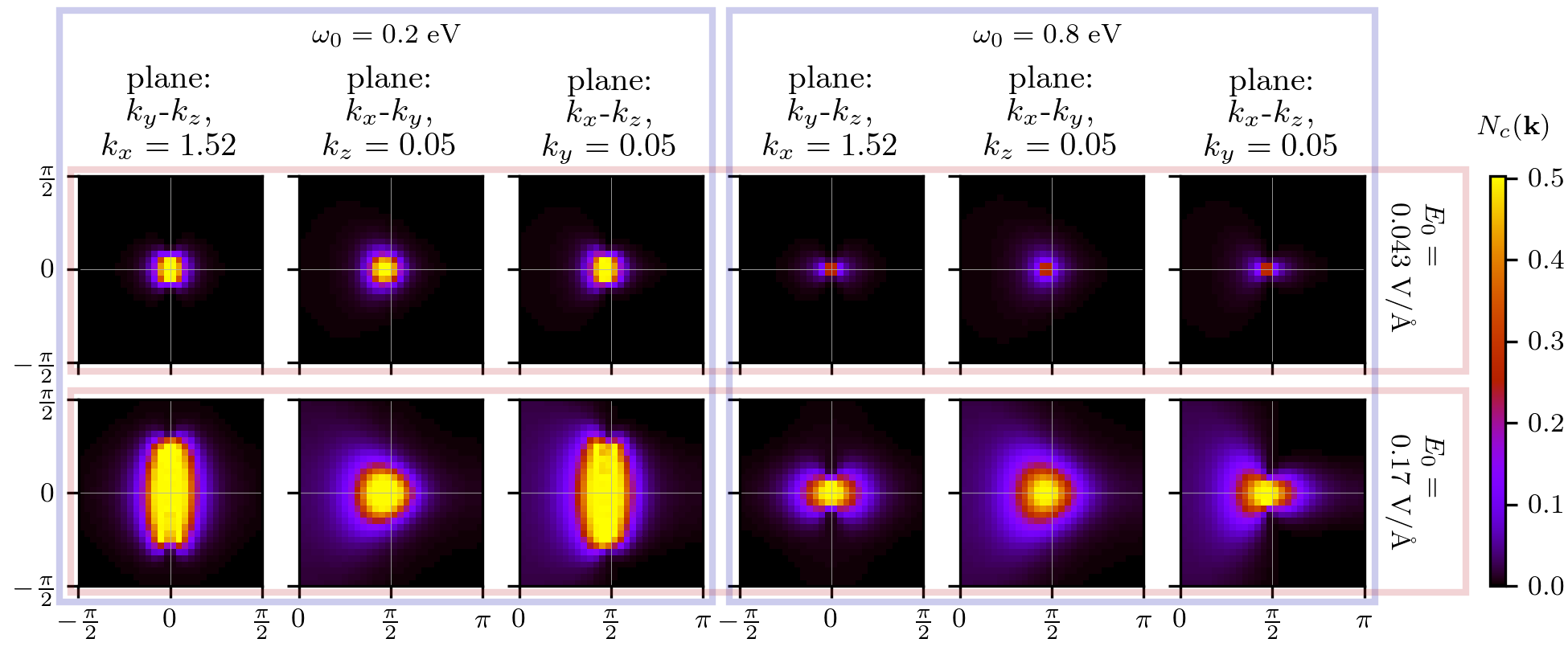}
\caption{Excitations $N_c(\mathbf{k})$ at $t=T/2$ for two different values of $\omega_0$ and $E_0$. Excitations are presented as cross sections ($k_y$–$k_z$, $k_x$–$k_y$, and $k_x$–$k_z$), traversing near the Weyl node at $\mathbf{k}_0 = (1.47,0,0)$. The parameter $\gamma$ is set to $2.1$.}
\label{fig:excitationsDifferentE}
\end{figure}

When $E_0$ is increased, the total magnitude of the response is increased, correlating with the density of the excitations (up to the saturation point $N_c=0.5$) and their extent in the BZ that both increase (see Fig. \ref{fig:excitationsDifferentE}). The response increases also in the strict Weyl regime around $\gamma=2$. This is rationalized by a weakening of the well-separatedness criterion as seen in the insets of Fig.~\ref{fig:figHHGgammaDifferentE} with contributions to the third harmonic between the two Weyl nodes that become more significant with increasing $E_0$. 

The dependence on $\omega_0$ is more subtle. With progressively increasing $\omega_0$ the magnitude of response in Fig. \ref{fig:figHHGgammaDifferentE} drops and the $\gamma$ dependence becomes significantly milder. We rationalize this by noting that as $\omega_0$ increases, the resonant condition $\hbar\omega_{cv} = n \hbar\omega_0$ ($n \in \mathbb{N}$)~\citep{avetissian2012creation} shifts further away from the nodes. The transition dipole moments $\mathbf{d}_{vc}$ decrease there~\citep{avetissian2022high}. As a result, for large $\omega_0$, the density of excitations reduces near the Weyl nodes 
and increases somewhat away from them (see Fig. \ref{fig:excitationsDifferentE}). Thus, with increasing $\omega_0$ one is less sensitive to the low-energy details, and the response as a function of $\gamma$ becomes more uniform.  [We note that when changing $\omega_0$, we have also scaled all the other temporal scales ($T$, $\tau$, and $T_2$) such that the characteristics of the pulse remain unchanged.]

\begin{figure}
\centering
\includegraphics[width=.95\textwidth]{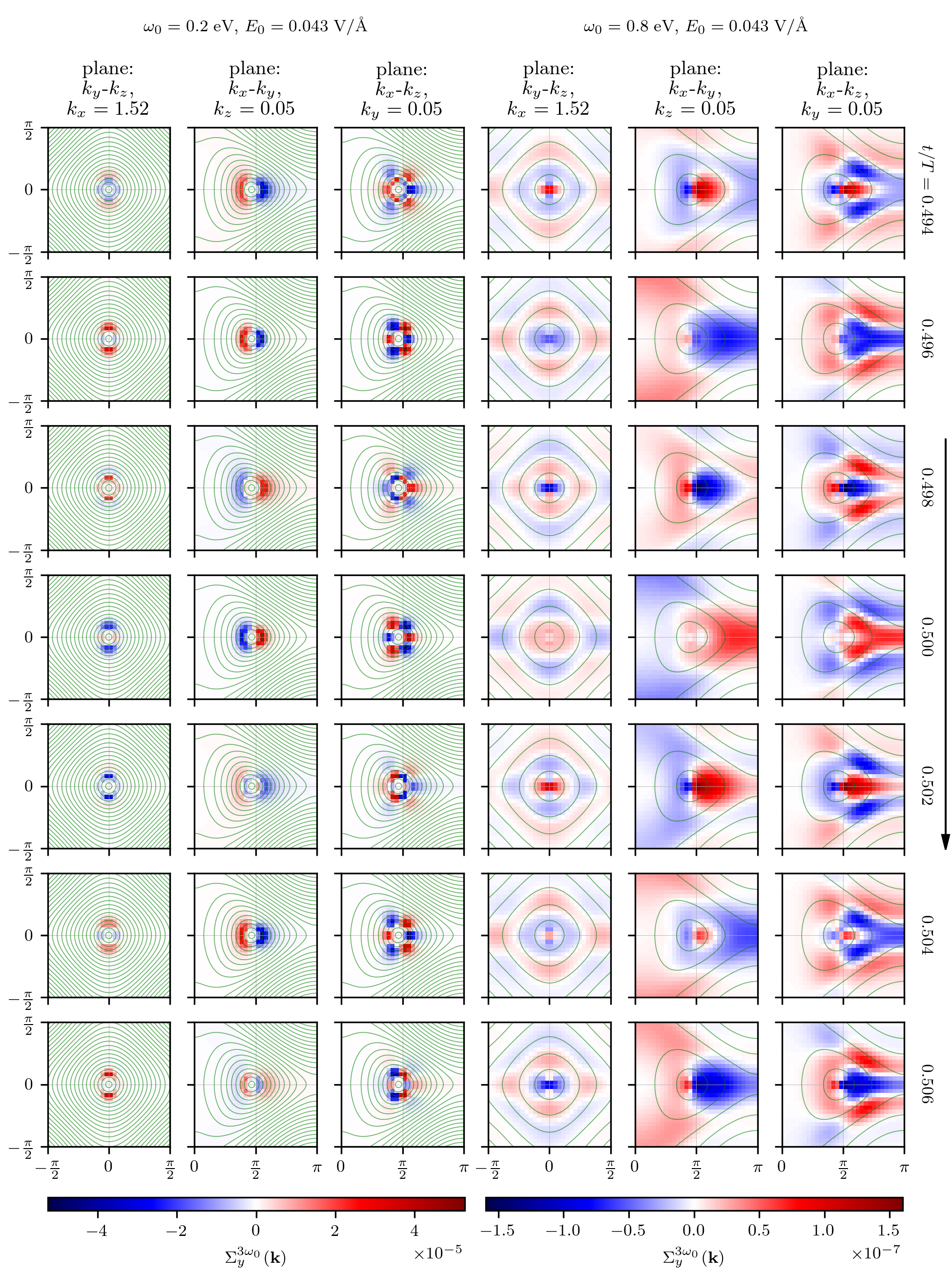}
\caption{Time evolution of the third harmonic ($\omega=3\omega_0$) of the combined quantity $\Sigma_y(\mathbf{k}) = (J_y(k_x,k_y,k_z) + J_y(k_x,k_y,-k_z))/2$. The contours indicate the resonant conditions $\hbar\omega_{cv} = n \hbar\omega_0$ for $n=1,2,3\dots$}
\label{fig:figTimeEvolutionJ3}
\end{figure}

To illustrate how the response is distributed in the BZ, and to illustrate how localization of the response to the Weyl node is affected by $\omega_0$, we show maps of the time evolution of the third harmonic contribution to the current on Fig.~\ref{fig:figTimeEvolutionJ3} at $\gamma=2.1$. Because we found that the current is nearly odd when $k_z$ is flipped $J_y(k_x,k_y,k_z) \simeq -J_y(k_x,k_y,-k_z)$, we show combined quantity $\Sigma_y(\mathbf{k}) = (J_y(k_x,k_y,k_z) + J_y(k_x,k_y,-k_z))/2$. These time evolutions are displayed as cross sections ($k_y$–$k_z$, $k_x$–$k_y$, and $k_x$–$k_z$), traversing near the Weyl node at $\mathbf{k}_0 = (1.47,0,0)$. Examining these results, we observe that the contributions to higher harmonics are concentrated near the Weyl node in the well-separated regime ($\omega_0=0.2$ eV). However, for $\omega_0=0.8$ eV, these contributions are present even at $\mathbf{k}$ far from $\mathbf{k}_0$, again indicating that the assumption of well-separated nodes does not apply for large $\omega_0$.

\begin{figure}
\centering
\includegraphics[width=.95\textwidth]{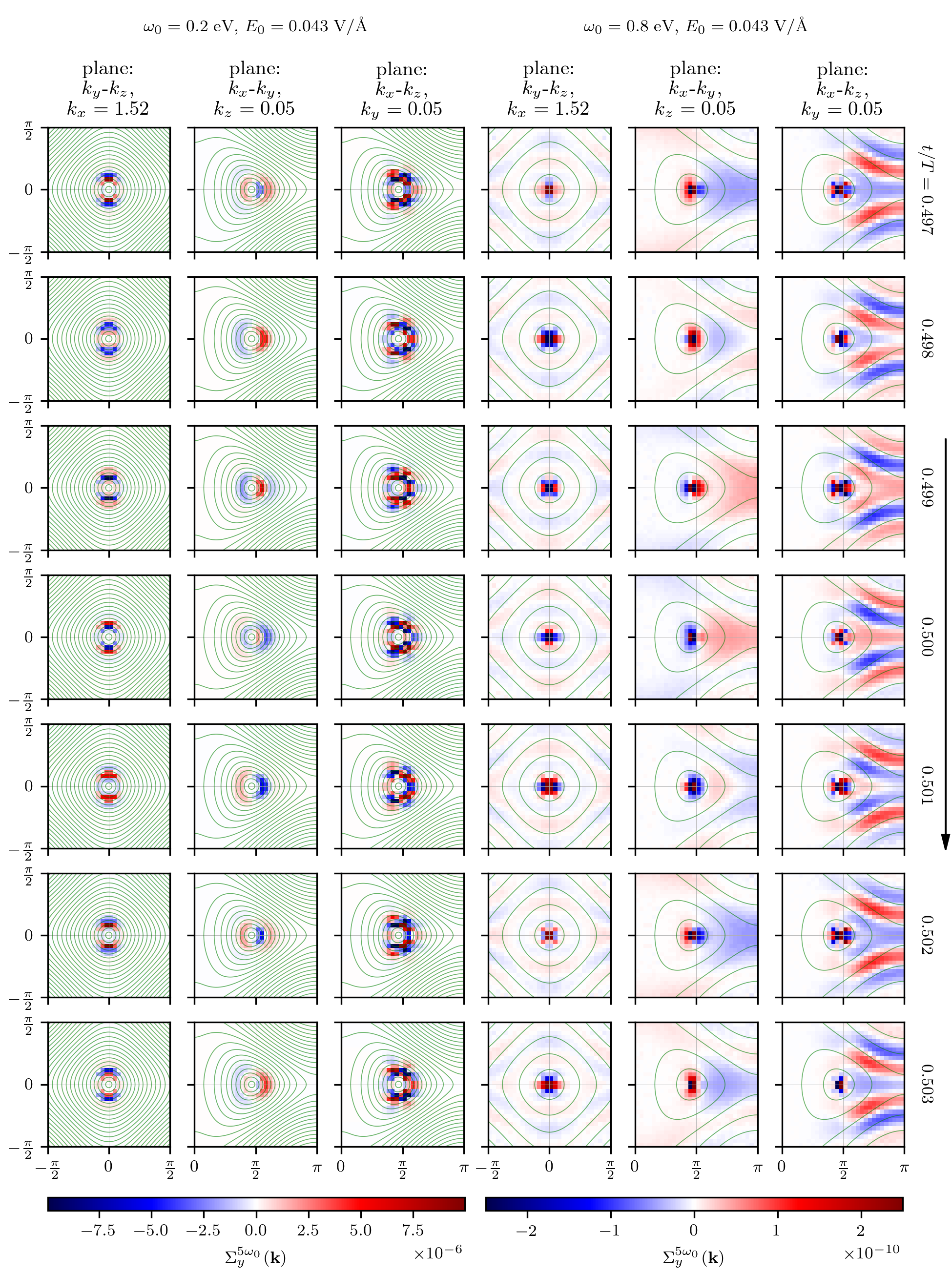}
\caption{Time evolution of the fifth harmonic ($\omega=5\omega_0$) of the combined quantity $\Sigma_y(\mathbf{k}) = (J_y(k_x,k_y,k_z) + J_y(k_x,k_y,-k_z))/2$. The contours indicate the resonant conditions $\hbar\omega_{cv} = n \hbar\omega_0$ for $n=1,2,3\dots$}
\label{fig:figTimeEvolutionJ5}
\end{figure}

The observations on the localization of contributions, when $\omega_0$ is small, can also be made for higher harmonics. The fifth harmonic is shown in Fig. \ref{fig:figTimeEvolutionJ5}. Although oscillations are faster compared to the third harmonic (note that the snapshots are taken at shorter time intervals as in Fig. \ref{fig:figTimeEvolutionJ3}). As expected, the frequencies of these oscillations are $\omega = m \omega_0$, where $m$ denotes the harmonic order, specifically 3 and 5.

\section{S4. $C_{2z}$ rotational symmetry and vanishing anomalous current for a well-separated Weyl node}
\label{appendix:iv}

In this section, we investigate the behavior of a well-separated Weyl node characterized by linear dispersion in the low-energy limit, following the approach outlined in Ref. \cite{avetissian2022high}. The Hamiltonian for an anisotropic Weyl cone is given by $H(\mathbf{q}) = \sum_i v_i q_i \sigma_i$, where $v_i$ are the Fermi velocities, $\mathbf{q}=(q_x,q_y,q_z)^T$ is the displacement vector from the Weyl node, and $\sigma_i$ are Pauli matrices. Suppose we apply an electric field in the $z$ direction. In this case, the Hamiltonian preserves the $C_{2z}$ rotational symmetry in the $q_x$--$q_y$ plane, given by:
\begin{equation}
    \sigma_z H(\mathbf{q}(t)) \sigma_z = H(\mathbf{q}'(t)), \tag{S4.1}
\end{equation}
where we use the notation $\mathbf{q}(t) = \mathbf{q} + \mathbf{A}(t)$, $\mathbf{q}'(t) = \mathbf{q}' + \mathbf{A}(t)$, and $\mathbf{q}' = (-q_x,-q_y,q_z)$.

Now let us derive the anomalous current produced by a well-separated node. Using the notation $N_{c/v} = \rho_{cc/vv}$ and $P = \rho_{vc} = \rho_{cv}^*$ (omitting superscripts $\mathbf{q}$ for brevity), we rewrite the SBE, Eq. (2) of the main text, as:
\begin{align}
    \partial_t N_c(t) &= -\partial_t N_v(t) = 2 \mathbf{E}(t) \cdot\mathfrak{Im}\!\left\{ \mathbf{d}_{cv} P(t) \right\} \tag{S4.2}\\
    \partial_t P(t) &= \left[i \omega_{cv} - 1/T_2 +i \mathbf{E}(t) \cdot \left(\mathbf{d}_{cc}-\mathbf{d}_{vv}\right)\right] P(t) \nonumber \\
    &+ i \mathbf{E}(t) \cdot \mathbf{d}_{cv}^* \left(1-2 N_c(t)\right). \tag{S4.3}
\end{align}

For $z$ polarization of the electric field, we have $(\mathbf{d}_{cc} - \mathbf{d}_{vv})_z = 0$ (see Ref. \citep{avetissian2022high}). Now, let us consider the transformation $\mathbf{q} \to \mathbf{q}'$ which leads to transformations:
\begin{align}
    E_z &\to E_z \tag{S4.4}\\
    \omega_{mn} &\to \omega_{mn} \tag{S4.5}\\
\label{eq:apx3:d_y}
    (\mathbf{d}_{cv})_y &\to (\mathbf{d}_{cv})_y \tag{S4.6}\\
    (\mathbf{d}_{cv})_z &\to -(\mathbf{d}_{cv})_z. \tag{S4.7}
\end{align}
Combining this with the initial conditions $N_v(0) = 1$, $N_c(0) = P(0) = 0$, we conclude that the density matrix elements transform as $N_{c/v} \to N_{c/v}$ and $P \to -P$.

Next, we rewrite the contributions to the anomalous current ${J}_y(\mathbf{q})$ as:
\begin{equation}
    {J}_y(\mathbf{q}) = -\sum_m (\mathbf{V}_m)_y N_m + 2 \omega_{cv} \mathfrak{Im}\!\left\{ (\mathbf{d}_{cv})_y P \right\} \tag{S4.8}
\end{equation}
where $\mathbf{V}_c = -\mathbf{V}_v = \mathbf{p}_{cc} = \nabla_\mathbf{q} \epsilon_c = \mathbf{q} / \epsilon_c$ is evaluated near the Weyl node, where the dispersion is linear. The $y$ component of $\mathbf{V}_c$ transforms as $(\mathbf{V}_c)_y \to -(\mathbf{V}_c)_y$. Together with Eq. (\ref{eq:apx3:d_y}) and the transformations for $N_{c/v}$ and $P$, we find ${J}_y(\mathbf{q}) = -{J}_y(\mathbf{q}')$. This implies that the contributions to the total anomalous current cancel in pairs for each well-separated Weyl node:
\begin{align}
    {J}_y &= \sum_\mathbf{q} {J}_y(\mathbf{q}) = \frac{1}{2} \left(\sum_\mathbf{q} {J}_y(\mathbf{q}) + \sum_\mathbf{q'} {J}_y(\mathbf{q'})\right) \nonumber \\
    &= \frac{1}{2}\sum_\mathbf{q} \left({J}_y(\mathbf{q}) + {J}_y(\mathbf{q'})\right) = 0. \tag{S4.9}
\end{align}

\section{S5. Tilted cones in WSM}
\label{appendix:v}
In this section, we introduce an additional term in the Hamiltonian to describe the tilt of the Weyl cones. The modified Hamiltonian, denoted as $H'(\mathbf{k})$, is given by
\begin{equation}
H'(\mathbf{k}) = d_0(\mathbf{k}) I + H(\mathbf{k}), \tag{S5.1}
\end{equation}
where $H(\mathbf{k})$ is the original Hamiltonian, and $I$ is the $2 \times 2$ identity matrix. The function $d_0(\mathbf{k})$ represents an arbitrary periodic function of the crystal momentum $\mathbf{k}$. We choose this function in a way that it tilts the Weyl cones while maintaining the occupancy of the energy bands unchanged, resulting in a tilted type-I Weyl semimetal. A specific example of such a function is $d_0 = \delta (\cos(k_x) - \gamma)$, where $\delta$ is a parameter that quantifies the tilt along the $k_x$ direction.

The introduction of $d_0(\mathbf{k})$ causes a shift in the energy spectrum, given by $\epsilon'_{c,v}(\mathbf{k}) = d_0(\mathbf{k}) + \epsilon_{c,v}(\mathbf{k})$. However, despite this modification, the corresponding wavefunctions remain unchanged: $|n,\mathbf{k}\rangle' = |n,\mathbf{k}\rangle$. Additionally, the transition frequencies and transition dipole moments are also unaffected, namely $\omega'^{\mathbf{k}}_{mn} = \omega_{mn}^{\mathbf{k}}$ and $\mathbf{d}'^{\mathbf{k}}_{mn} = \mathbf{d}_{mn}^{\mathbf{k}}$. Furthermore, the off-diagonal elements ($m \neq n$) of the group-velocity matrix are also unaffected:
\begin{equation}
\label{eq:appendix:p_off-diag}
    \mathbf{p}'^{\mathbf{k}}_{mn} = i \mathbf{d}'^{\mathbf{k}}_{mn} \omega'^{\mathbf{k}}_{mn} = i \mathbf{d}^{\mathbf{k}}_{mn} \omega^{\mathbf{k}}_{mn} = \mathbf{p}^{\mathbf{k}}_{mn}, \tag{S5.2} 
\end{equation}
however, the diagonal elements ($m = n$) change as follows:
\begin{equation}
\label{eq:appendix:p_diag}
\mathbf{p}'^{\mathbf{k}}_{mm} = \mathbf{V}'^{\mathbf{k}}_{m} = \nabla_\mathbf{k} \epsilon'^{\mathbf{k}}_m = \nabla_{\mathbf{k}} (d_0^{\mathbf{k}} + \epsilon^{\mathbf{k}}_m) = \nabla_{\mathbf{k}} d_0^{\mathbf{k}} + \mathbf{V}^{\mathbf{k}}_{m}. \tag{S5.3}
\end{equation}

When considering the time evolution of the density matrix (Eq. (2) of the main text) under the influence of the tilted Hamiltonian $H'(\mathbf{k})$, we find that if the initial conditions satisfy $\rho'^{\mathbf{k}}_{mn}(t=0) = \rho_{mn}^{\mathbf{k}}(t=0)$, which is true in the type-I Weyl semimetal, then the density matrix at any time $t$ remains the same for both the original and tilted Hamiltonians, $\rho'^{\mathbf{k}}_{mn}(t) = \rho_{mn}^{\mathbf{k}}(t)$.

Next, we consider the current density $\mathbf{J}'(t)$. By substituting the expressions (\ref{eq:appendix:p_off-diag}) and (\ref{eq:appendix:p_diag}) into the Eq. (3), we obtain the current density under the tilt as:
\begin{align}
\label{eq:appendix:J_tilt}
    \mathbf{J}'(t) &= - \sum_{\mathbf{k}}\left[\sum_m (\mathbf{V}'^{\mathbf{k}(t)}_m \rho'^{\mathbf{k}}_{mm}(t)) + \sum_{m \neq n} (\mathbf{p}'^{\mathbf{k}(t)}_{nm} \rho'^{\mathbf{k}}_{mn}(t))\right] = \mathbf{J}(t) - \sum_{\mathbf{k}} \left[\sum_m \nabla_{\mathbf{k}(t)} d^{\mathbf{k}(t)}_0 \rho_{mm}^{\mathbf{k}}(t)\right] \nonumber \\
    & \overset{\mathrm{(a)}}{=} \mathbf{J}(t) - \sum_{\mathbf{k}} \nabla_{\mathbf{k}} d^{\mathbf{k}}_0 \to \mathbf{J}(t) - 
    \int_{\mathrm{BZ}} \nabla_{\mathbf{k}}d_0^{\mathbf{k}}\, \mathrm{d}V_{\mathbf{k}} \overset{\mathrm{(b)}}{=} \mathbf{J}(t) - \int_{\partial \textrm{BZ}} d_0^{\mathbf{k}} \, \mathrm{d}\mathbf{S}_{\mathbf{k}} \overset{\mathrm{(c)}}{=} \mathbf{J}(t). \tag{S5.4}
\end{align}
In the above derivation, $\mathrm{BZ}$ denotes the Brillouin zone, and $\partial\mathrm{BZ}$ represents its boundary. Several relations have been employed in this process, including: (a) $\sum_m \rho_{mm}^{\mathbf{k}} = 1$, (b) the Gauss theorem, and (c) the periodicity of $d_{0}^\mathbf{k}$. Equation (\ref{eq:appendix:J_tilt}) illustrates that the tilt induced by $d_0$ does not alter the total current.

Note that the above argumentation relies on a single occupancy at every momentum point. However, this condition does not apply to over-tilted, type-II WSMs that exhibit hyperboloidal electron and hole pockets \citep{li2022high} where the occupancy is double or zero, respectively. We leave further investigation of type-II WSMs for future studies.

\section{S6. Decoupled left- and right-handed Weyl spinors}
\label{appendix:vi}

The linearized regime of well-separated Weyl nodes discussed in the main text is analogous to the model outlined in Ref. \citep{avetissian2022high}. In this reference, the authors explicitly treat two decoupled Hamiltonians, each corresponding to a distinct, left and right, chirality of the Weyl spinor. As mentioned in the main text, our findings differ from those reported in Ref. \citep{avetissian2022high}, specifically regarding the anomalous higher-order responses. The objective of this section is to clarify this matter.

It is crucial to note that the finite nature of the first harmonic stems from topologically non-trivial 2D slices, characterized by a non-zero Chern number, located between the Weyl nodes. Following Ref. \citep{avetissian2022high}, this corresponds to the fourth term in Eq. (22) which, upon integration, yields a linear response. However, the remaining terms in Eq. (22) of Ref. \citep{avetissian2022high} are multiplied by either $N_c$ (excitations) or $P$ (interband polarization), which invalidates the application of the same topological argument used for the first harmonic also to the higher ones.

\begin{figure}
    \centering
    \includegraphics[width=.6\textwidth]{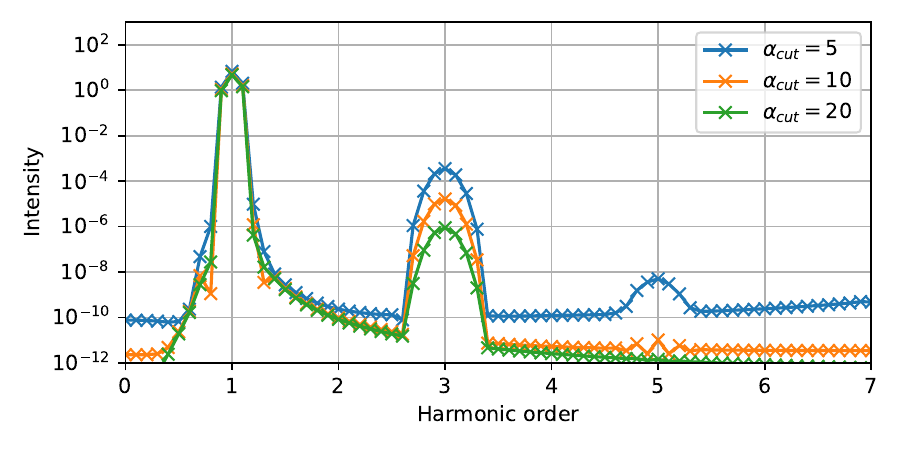}
    \caption{Results for the parameters of Ref. \citep{avetissian2022high} for various cut-off values $\alpha_{cut}$.}
    \label{fig:MkrtchianConvergence}
\end{figure}

To illustrate the different behavior between the first and higher harmonics in numerics, we reproduced the analysis presented in Fig. 3 of Ref. \citep{avetissian2022high}. Our results are displayed in Fig. \ref{fig:MkrtchianConvergence}, illustrating the anomalous harmonics using three different integration region cut-offs, $\alpha_{cut}$. The selected cut-offs offer evidence that higher harmonics diminish as $\alpha_{cut}$ is increased, whereas the first harmonic remains unchanged.

In the analytical treatment presented in Ref. \citep{avetissian2022high}, one has to carefully evaluate Eq. (23) (the fifth term of Eq. (22)), specifically when approximating $N_c(\textbf{k}_{0}, t)$. Despite the cancellation around Weyl nodes (as indicated by the $sgn$ function in Eq. (24) of Ref. \citep{avetissian2022high}), the substitution of $N_c(\textbf{k}_{0}, t)$ with $N_c(\textbf{k}_{0w}, t)$ renders contributions far from the Weyl nodes as finite, while excitations actually vanish there (as depicted in Fig. 1 of Ref. \citep{avetissian2022high}).

To resolve these issues, it is crucial to establish the rotational symmetry with respect to each isolated Weyl node, Eq. (6) of the main text, where contributions to anomalous higher-order harmonics cancel out in pairs. Specifically, this reduces the relevant integration interval in Eq. (24) of Ref. \citep{avetissian2022high} to $(\Lambda-2b, \Lambda)$:
\begin{align}
    j_y(t) &= \frac{2 E_z(t)}{(2\pi)^3} \int \textrm{d}^3 \mathbf{k}_0 \frac{\left(k_{0x}+b\right) N_c(\mathbf{k}_0, t)}{\left[(k_{0x}+b)^2 + k_{0y}^2 + k_{0z}^2\right]^{3/2}} \tag{S6.1} \\
    &= \frac{2 E_z(t)}{(2\pi)^2} \int_{\Lambda-2b}^{\Lambda} \textrm{d} k_{0x} \int_0^{\infty} k_\perp \textrm{d}k_\perp \frac{\left(k_{0x}+b\right) N_c(\mathbf{k}_0, t)}{\left[k_{\perp}^2+(k_{0x}+b)^2\right]^{3/2}}. \tag{S6.2}
\end{align}
Further, evaluating the upper bound of $j_y$ by using the substitution $N_c(\textbf{k}_{0}, t) \to N_c^{\textrm{max}}(t)$, where $N_c^{\textrm{max}}(t) = \max_{k_{0x} \in (\Lambda-2b, \Lambda)} \left[ N_c(\textbf{k}_{0}, t) \right]$, we get
\begin{align}
    j_y(t) &\leq \frac{2 E_z(t)}{(2\pi)^2} \int_{\Lambda-2b}^{\Lambda} \textrm{d} k_{0x} \int_0^{\infty} k_\perp \textrm{d}k_\perp \frac{\left(k_{0x}+b\right) N_c^{\textrm{max}}(t)}{\left[k_{\perp}^2+(k_{0x}+b)^2\right]^{3/2}} \tag{S6.3} \\
    &= \frac{b}{\pi^2} E_z(t) N_c^{\textrm{max}}(t) \to 0. \tag{S6.4}
\end{align}
In the final step we have considered the limit $\Lambda \to \infty$. The same reasoning applies to the first three terms in Eq. (22) of Ref. \citep{avetissian2022high}, demonstrating that the anomalous higher-order harmonics vanish in the limit $\Lambda \to \infty$ for the decoupled Weyl Hamiltonian.

Regarding Ref. \citep{avetissian2022high}, let us conclude with a comment on the significance of the separation between Weyl nodes and nonlinearities in the spectrum concerning the generation of anomalous higher-order harmonics. As established above, in the decoupled Weyl Hamiltonian, the anomalous higher-order response remains invariant, i.e. zero, under the separation between Weyl nodes. However, the introduction of nonlinearities leads to a finite anomalous higher-order response. Thus, in this simplified depiction, the importance of deviations from strict linearity for AHHG becomes clear.

\section{S7. Mirror symmetry}
\label{appendix:vii}

In this section, we introduce the notation $\mathbf{v}_\perp = (-v_x, v_y, v_z)$, where $\mathbf{v}$ represents an arbitrary 3D vector.

The Hamiltonian in the paper exhibits mirror symmetry: $H(\mathbf{k}) = H(\mathbf{k}_\perp)$. From this symmetry, we deduce:
\begin{equation}
    \mathbf{J}\left(\mathbf{k}_\perp, \mathbf{E}_\perp \right) = \mathbf{J}(\mathbf{k}, \mathbf{E})_\perp. \tag{S7.1}
\end{equation}

Consider the polarization in the $z$ direction. Then, $\mathbf{E}_\perp = \mathbf{E}$ holds, leading to:
\begin{equation}
    {J}_x = \sum_\mathbf{k} {J}_x(\mathbf{k},\mathbf{E}) = \sum_{k_x > 0} \left( {J}_x(\mathbf{k},\mathbf{E}) + {J}_x(\mathbf{k}_\perp,\mathbf{E}) \right) = 0, \tag{S7.2}
\end{equation}
i.e. there is no response in the direction of the Weyl node separation vector when the polarization of the laser is perpendicular to it.


%% file: bibOutput.bbl
%